\documentclass[twocolumn]{aastex631}
\usepackage{amsmath}
\usepackage{amssymb}
\usepackage{latexsym}
\usepackage{longtable}
\usepackage{verbatim}
\usepackage{natbib}
\usepackage{mathtools}
\usepackage{multirow}
\usepackage{empheq}
\usepackage{bm}
\usepackage{graphicx}
\usepackage{color}
\newcommand{\HH}{H$_2$}
\newcommand{\MHH}{$M_{\mathrm{mol}}$}

\newcommand{\HII}{H\,{\sc ii}}

\newcommand{\alphaCII}{$\alpha_{[{\mathrm{C}}\,\scriptsize{\textsc{ii}}]}$}

\newcommand{\OIIIFIR}{[O\,{\sc iii}]\,88\,$\mu$m}
\newcommand{\OIFIR}{[O\,{\sc i}]\,63\,$\mu$m}
\newcommand{\NII}{[N\,{\sc ii}]}
\newcommand{\CII}{[C\,{\sc ii}]}
\newcommand{\CI}{[C\,{\sc i}]}
\newcommand{\COI}{CO(1$-$0)}

\newcommand{\LNII}{$L_{\rm [N\,{\scriptsize \textsc{ii}}]205\,\mu{\rm m}}$}
\newcommand{\LNIIbc}{$L_{\rm [N\,{\scriptsize \textsc{ii}}]122\,\mu{\rm m}}$}

\newcommand{\LIR}{$L_{\rm IR}$}
\newcommand{\LFIR}{$L_{\rm FIR}$}
\newcommand{\LCO}{$L_{\rm CO(1-0)}$}
\newcommand{\LOIII}{$L_{\rm [OIII]88\,\mu{\rm m}}$}
\newcommand{\LCII}{$L_{\rm [C\,{\scriptsize \textsc{ii}}]}$}

\newcommand{\LNIIab}{$L_{\rm [N\,{\scriptsize \textsc{ii}}]205\,\mu {\rm m}}$}

\newcommand{\NIIab}{{\rm [N\,{\sc ii}]$\,{205\,\mu {\rm m}}$}}
\newcommand{\NIIbc}{{\rm [N\,{\sc ii}]$\,{122\,\mu {\rm m}}$}}

\newcommand{\mum}{$\mu$m}
\newcommand{\herschel}{{\it Herschel}}

\newcommand{\kkmspc}{K~km~s$^{-1}$~pc$^2$}

\shorttitle{Ionized Carbon Emission in Galaxies}
\shortauthors{Zhao et al.}

\begin{document}
\defcitealias{2018MNRAS.481.1976Z}{Z18}

\title{Ionized Carbon in Galaxies: The \CII\ 158\,\mum\ Line as a Total Molecular Gas Mass Tracer Revisited\footnote{Released on XX, XX, XXXX}}

\correspondingauthor{Yinghe Zhao}
\email{zhaoyinghe@ynao.ac.cn}

\author{Yinghe Zhao}
\affiliation{Yunnan Observatories, Chinese Academy of Sciences, Kunming 652016, China}
\affiliation{Key Laboratory of Radio Astronomy and Technology (Chinese Academy of Sciences), A20 Datun Road, Chaoyang District, Beijing, 100101, P. R. China}

\author{Jiamin Liu}
\affiliation{Yunnan Observatories, Chinese Academy of Sciences, Kunming 652016, China}
\affiliation{University of Chinese Academy of Sciences, Beijing 100049, China}

\author{Zhi-Yu Zhang}
\affiliation{School of Astronomy and Space Science, Nanjing University, Nanjing 210093, China}
\affiliation{Key Laboratory of Modern Astronomy and Astrophysics, Nanjing University, Ministry of Education, Nanjing 210093, China}

\author{Thomas G. Bisbas}
\affiliation{Research Center for Astronomical Computing, Zhejiang Lab, Hangzhou 311100, China}



\begin{abstract}
In this paper we present a statistical study of the \CII\ 158 \mum\ line and the \COI\ emission for a sample of $\sim$200 local and high-$z$ (32 sources with $z>1$) galaxies with much different physical conditions. We explore the correlation between the luminosities of \CII\ and \COI\ lines, and obtain a strong linear relationship, confirming that \CII\ is able to trace total molecular gas mass, with a small difference between (U)LIRGs and less-luminous galaxies. The tight and linear relation between \CII\ and \COI\ is likely determined by the average value of the observed visual extinction $A_V$ and the range of $G_0/n$ in galaxies. Further investigations into the dependence of \LCII/\LCO\ on different physical properties show that \LCII/\LCO\ (1) anti-correlates with $\Sigma_{\mathrm{IR}}$, and the correlation becomes steeper when $\Sigma_{\mathrm{IR}} \gtrsim 10^{11}$ $L_\odot\,\mathrm{kpc}^{-2}$; (2) correlates positively with the distance from the main sequence $\Delta(\mathrm{MS})$ when $\Delta(\mathrm{MS})\lesssim0$; and (3) tends to show a systematically smaller value in systems where the \CII\ emission is dominated by ionized gas. Our results imply that caution needs to be taken when applying a constant \CII-to-\MHH\ conversion factor to estimate the molecular gas content in extreme cases, such as galaxies having low-level star formation activity or high SFR surface density. 
\end{abstract}



\section{Introduction} \label{sec:intro}
Star formation (SF) converts gas into stars, and thus is one of the most fundamental drivers of galaxy evolution. To better understand the relation between SF and molecular gas, it is crucial to know the molecular gas content, which has usually been traced by the low-$J$ transitions of carbon monoxide (CO; e.g., \citealt{2005ARA&A..43..677S, 2013ARA&A..51..207B}), in external galaxies. However, it becomes challenging to detect the low-$J$ CO transitions in higher redshift galaxies due to (1) the limited sensitivities of currently available ground facilities; (2) the combination of low excitation temperature of \COI\ and  increasing cosmic microwave background (e.g., \citealt{2016RSOS....360025Z}), reducing the observable brightness temperature contrast of the line (e.g., \citealt{2020ApJ...895...81R}); and (3) the decreasing metal content, causing weaker CO emission (and it would no longer trace \HH\ gas under extremely metal-poor conditions; e.g., \citealt{2015ApJ...804L..11S}). Therefore, it is important to have some alternative \HH\ tracers, as more and more galaxies have been identified at high redshifts, with some quasars and other emission-line galaxies discovered at extremely high redshifts of $z > 7$  (e.g., \citealt{2013Natur.502..524F,2015Natur.519..327W,2017ApJ...845..154V,2018Natur.553..473B,2022ApJ...931..160B}).

Recently, the \CI\ emission (\CI\ $^3{\mathrm{P}}_1\rightarrow$~$^3{\mathrm{P}}_0$ at 492.161 GHz, hereafter \CI(1$–$0); and $^3{\mathrm{P}}_2\rightarrow$~$^3{\mathrm{P}}_1$ at 809.344 GHz, hereafter \CI(2$–$1)) has been proposed and observationally confirmed its potential as a total molecular gas mass (\MHH) tracer (e.g. \citealt{2004ApJ...615L..29P, 2011ApJ...730...18W, 2014ApJ...797L..17L, 2014MNRAS.440L..81O, 2017ApJ...840L..18J, 2019ApJ...880..133J, 2018ApJ...869...27V, 2020ApJ...890...24V,2022MNRAS.517..962D}). However, the \CI\ lines are generally faint and thus they are more difficult to detect at high redshifts. For $z\ga 4$, there are only about twenty lensed sources detected in [C\,{\sc i}](1$-$0) and/or [C\,{\sc i}](2$-$1), and most of which are marginally detected with a signal-to-noise ratio (S/N) of $3-5$ (and specially true for the [C\,{\sc i}](1$-$0) line; \citealt{2022MNRAS.511.3017U, 2023A&A...676A..89G} and references therein; \citealt{2023MNRAS.521.5508H}). Furthermore, low [C\,{\sc i}](1$-$0)-to-CO ratio has also been found in local luminous infrared galaxies (LIRGs; $L_{\rm IR}$[8$-$1000\,$\mu$m]~$>10^{11}$~$L_\odot$; \citealt{1996ARA&A..34..749S}) such as NGC~6052 and NGC~7679 (\citealt{2020ApJ...897L..19M,2021ApJS..257...28M}), and in some strongly lensed sources at high-$z$ (\citealt{2021ApJ...908...95H}), which might be due to the subthermal excitation (\citealt{2022MNRAS.510..725P}) and/or to the presence of an $\alpha$-enhanced ISM (\citealt{2024MNRAS.527.8886B}), and thus challenge its capability of tracing total molecular gas mass. 

Alternatively, the \CII~$^2{\mathrm{P}}_{3/2} \rightarrow$~$^2{\mathrm{P}}_{1/2}$ fine-structure transition at 158 \mum\ (1900.5~GHz; hereafter \CII) might be a promising tool to probe the gas content in the distant Universe (e.g. \citealt{2013ARA&A..51..105C}; \citealp[][herealfter Z18]{2018MNRAS.481.1976Z}; \citealt{2020A&A...643A...5D,2020A&A...643A.141M,2021ApJ...922..147H}; \citealt{2021MNRAS.503.4878S,2022ApJ...929...92V}), due to the fact that (1) It can arise from all the three phases of gas, being able to be excited via collisions with electrons, atomic and molecular hydrogen with relatively low critical densities (e.g., \citealt{2012ApJS..203...13G}); (2) The \CII\ emission is a dominant coolant of the neutral gas and in general the most luminous single far-infrared (FIR) line in star-forming galaxies, accounting for $0.1\%-1\%$ of the total FIR luminosity (e.g., \citealt{1991ApJ...373..423S,2001ApJ...561..766M,2017ApJ...846...32D}), and thus making it easy to detect in high-$z$ sources (e.g., \citealp[$z=8.31$]{2020MNRAS.493.4294B}; \citealp[a low-mass galaxy at $z\sim6.3$]{2023arXiv230911548G}; \citealp[$z=7.31$]{2023MNRAS.524.1775H}; \citealp[$z\sim7$]{2023ApJ...954..103S}); 
and/or (3) The dust continuum, another alternative to CO and a widely used tracer of \MHH\ (e.g. \citealt{2010A&A...518L.154S,2012ApJ...760....6M,2014ApJ...783...84S,2016ApJ...820...83S}), is fainter at higher redshifts due to cosmic dimming, lower metallicities and radiative transfer with the CMB (\citealt{2013ApJ...766...13D,2016RSOS....360025Z}). Indeed, sources with strong FIR line emission but weak dust continuum have been found at $z\sim2.5$  (\citealt{2015ApJ...806..260F}) and $z\sim5$ (\citealt{2014ApJ...782L..17D,2015Natur.522..455C}).  

However, it is necessary to validate the reliability of the \CII\ emission as a total {\it molecular} gas mass tracer due to the following reasons. Firstly, the gas masses in \citetalias{2018MNRAS.481.1976Z} are obtained with different methods (i.e., conversion from CO luminosity and Kennicutt-Schmidt (KS hereafter) relation in \citealt{2014ApJ...793...19S}), which might introduce bias to some extent. Secondly, as already noted in \citetalias{2018MNRAS.481.1976Z}, their sample galaxies with available \MHH\ estimates (see the top panel in their Figure 8) are dominated by main-sequence systems, and include few local extreme SF galaxies (e.g., ULIRGs) and low-SFE objects (e.g., early-type galaxies, hereafter ETGs; please see the proposed application of \CII\ in distant quiescent sources in \citealt{2023A&A...678L...9D}). The \CII\ flux in these two types of sources might be dominated by different gas phases (e.g., neutral gas from photodissociation regions (PDRs) vs diffuse ionized gas; \citealt{2015ApJ...802L..11L,2017ApJ...845...96C}) because of their much different physical environments (for example, \CII\ might be saturate at high SFR surface density; e.g., \citealt{2016MNRAS.463.2085M,2019MNRAS.489....1F,2022ApJ...934..115B}). Thirdly, the relative carbon-to-oxygen abundance ratio, an important ISM parameter affecting the \CII-to-\COI\ flux ratio as revealed by the latest astrochemical modelling (\citealt{2024MNRAS.527.8886B}), may be varied in different redshifts and/or different types of galaxies (such as discussed in \citealt{2018Natur.558..260Z}). Therefore, a more consistent way, using a heterogeneous data set covering a larger parameter space, is needed to check the capability of the \CII\ emission as a tracer of \MHH.
 
In this paper we present our comparison between the \COI\ and \CII\ emission for a large sample ($\sim$200 members) of local and high-$z$ galaxies, including gas-rich ETGs, main-sequence sources, and (U)LIRGs. Combining with other FIR lines such \NII\ and \OIIIFIR, it also allows us to investigate the properties of \CII\ emission under different physical conditions and thus to further explore its reliability of tracing the total molecular gas in galaxies. The remainder of this paper is organized as follows: we give a brief introduction of the sample, observations and data reduction in Section 2, present the results and discussion in Section 3, and briefly summarize the main conclusions in the last section. Throughout the paper, we adopt a Hubble constant of $H_0=70~$km~s$^{-1}$~Mpc$^{-1}$, $\Omega_{\rm M} =0.28$, and $\Omega_\Lambda=0.72$, which are based on the five-year WMAP results (\citealt{2009ApJS..180..225H}), and are the same as those used by the Great Observatories All-Sky LIRGs Survey (GOALS; \citealt{2009PASP..121..559A}).

\section{Sample and Data}{\label{sec:sample}}
\subsection{Main Sample}
Our approach cross-calibrates \CII\ as a gas mass tracer through a comparison with \COI\ emission, utilizing the feature that the \COI\ line can trace the total molecular gas in galaxies, and thus it relies on the gas mass obtained with \COI-to-\HH\ conversion factor $\alpha_{\mathrm{CO}}$. As discussed in \citet[and references therein]{2021MNRAS.502.2701B}, $\alpha_{\mathrm{CO}}$ is more sensitive to metallicity ($Z$) rather than the FUV intensity or the cosmic ray ionization rate  and is likely to have an upturn with decreasing metallicity below $1/3\sim1/2~Z_\odot$ (\citealt{2013ARA&A..51..207B}). Furthermore, the metallicity-dependent $\alpha_{\mathrm{CO}}$ obtained by different works show a much larger spread when $Z \lesssim 0.5Z_\odot$ (e.g., see Figure 10b in \citealt{2020A&A...643A.141M}). Therefore, we select sources (1) having both CO and \CII\ data available in the literature, and (2) likely being metal-rich ($Z\gtrsim0.5~Z\odot$) according to the stellar mass-metallicity (gas phase) relations for local (\citealt{2004ApJ...613..898T}; i.e., $Z \sim Z_\odot$ at stellar mass $M_\star$ of $\sim10^{9.2}~M_\odot$) and high-$z$ (\citealt[and references therein]{2023ApJ...942...24S}; i.e., $Z \gtrsim 0.5 Z_\odot$ for $M_\star\sim 10^{9-10}~M_\odot$ at $z=0.8-3.3$) galaxies. In order to explore the \CII's capability of tracing the total molecular gas mass under different physical conditions, we include different types of galaxies at various redshifts. We briefly describe the properties of different subsample in the following.

(1) {\it Local ETGs} (\citealt{2017ApJ...840...51L}). This sample includes 20 massive early-type sources with stellar mass of $M_\star \sim 10^{9.7}-10^{11}~M_\odot$, which are estimated using the $K$-band absolute magnitudes given in \cite{2011MNRAS.413..813C} and a typical mass-to-light ratio of $M_\star/L_K = 0.6$ in solar units (\citealt{2003ApJS..149..289B}) for a Kroupa initial mass function (IMF; \citealt{2002Sci...295...82K}). The star formation rates (SFRs) are in the range of $\sim0.01 - {\mathrm {several}}$~ $M_\odot$~yr$^{-1}$ (\citealt{2014MNRAS.444.3427D}). They were observed with the \herschel/PACS, and a part were also observed with the \herschel/SPIRE. The \CII\ data are adopted from \cite{2017ApJ...840...51L}, and CO data are adopted from \citet{2009AJ....137.3053Y}, \cite{2011MNRAS.410.1197C} and \cite{2013MNRAS.432.1796A}. We also collect metallicity measurements (\citealp{2018MNRAS.475.1485M,2019A&A...623A...5D,2019MNRAS.484..562G}) for six sources. In addition, we derive the metallicities, $12+\log ({\mathrm{O/H}}) = 8.68$ and 8.83 respectively for IC1024 and UGC06176, using the emission line data from \cite{2023A&A...671A.118C} and the O3N2 estimator from \citet[hereafter PP04]{2004MNRAS.348L..59P}.

(2) {\it Main sequence (MS) galaxies at $z\sim0.01-0.02$} (\citealt{2017MNRAS.470.4750A}). This sample is contains 23 objects with low-to-intermediate stellar mass ($10^9 - 10^{10}~M_\odot$), and with SFRs in the range of $\sim0.1-5~M_\odot$~yr$^{-1}$. The gas-phase metallicities are larger than $\sim$0.5 $Z_\odot$ (\citealt{2017MNRAS.470.4750A}). They have \herschel/PACS \CII\ and IRAM \COI\ observations.

(3) {\it H-ATLAS survey, with galaxies at $z=0.02-0.2$} (\citealt{2015MNRAS.449.2498I}). This sample of galaxies is a mixture of normal star-forming sources and LIRGs, with $L_{\mathrm{IR}} \sim 10^{10}-10^{12}~L_\odot$ and $M_\star \sim 10^{10}-10^{11}~M_\odot$, and is selected from the Herschel-Astrophysical Terahertz Large Area Survey (H-ATLAS; \citealt{2010PASP..122..499E}). They were observed with \herschel/PACS for the \CII\ emission (\citealt{2015MNRAS.449.2498I}) and with ALMA for the \COI\ line (\citealt{2017MNRAS.470.3775V}; also see \citealt{2017A&A...602A..49H}). In total 24 objects have both \CII\ and \COI\ detections, and 9 out of which have their metallicity available (\citealt{2015MNRAS.449.2498I}). 

(4) {\it GOALS LIRGs} (\citealt{2017ApJ...846...32D}). This sample of galaxies includes 75 sources with both \CII\ and \COI\ data available, including 65 LIRGs and 10 ULIRGs with $L_{\mathrm{IR}} > 10^{12}~L_\odot$ and $M_\star \sim 10^{10}-10^{12}~M_\odot$ (\citealt{2010ApJ...715..572H,2012ApJS..203....9U}), observed with the \herschel\ PACS and SPIRE. The \COI\ data for 71 out of these sources are collected by (\citealp[and references therein]{2017ApJ...840L..18J}), and for the rest 4 sources are adopted from \cite{1990A&A...236..327M}, \cite{1996A&AS..116..193C}, \cite{2017ApJ...844...96Y} and \cite{2023A&A...673A..13M}. For 13 sources the metallicities are adopted from \cite{2024NatAs...8..368P}, who derived the chemical abundances using fine-structure lines in the mid- to far-infrared range (\citealt{2021A&A...652A..23F}). To match the abundances measured with the PP04 calibration, we add $\sim$0.1 dex, which is obtained by comparing the metallicities of about 30 sources measured with both methods (Y. Zhao et al. 2024, in preparation), to the IR-based values. For additional 35 objects, we calculate their metallicities using the integrated line fluxes presented in \cite{2006ApJS..164...81M} and the PP04 calibration.

(5) {\it Local ULIRGs} (\citealt{2013ApJ...776...38F,2018ApJ...861...94H}). This sample of galaxies, which represent the most extreme starburst/AGN systems in the local Universe, include 21 members having both \CII\ and \COI\ observations. There is no systematic determination of stellar mass and metallicity for this sample objects since it is complicated by contributions from AGN and by the range of obscurations. Nevertheless, \cite{2011ApJ...732...72H} find that ULIRGs have stellar masses of $10^{10}–10^{12}~M_\odot$, and \cite{2008ApJ...674..172R} and \cite{2014ApJ...797...54K} find that local ULIRGs have a nearly Solar metallicity ($12+\log \mathrm{(O/H)}=8.69$; \citealt{2009ARA&A..47..481A}). The \COI\ data are collected from \cite{1990A&A...236..327M}, \cite{1992A&A...265..429A}, \cite{1997ApJ...478..144S}, \cite{2008A&A...477..747B}, \cite{2012ApJ...750...92X} and \cite{2023A&A...673A..13M}, and the metallicities for 3 sources are taken from \cite{2008ApJ...674..172R} and \cite{2014ApJ...797...54K}, which have been converted to the PP04 calibration using the correlation provided in \cite{2008ApJ...681.1183K}.

(6) {\it Lyman-break analogs (LBAs) at $z\sim0.2$} (\citealt{2017A&A...606A..86C}). This sample of LBAs, which are UV luminous ($L_{\mathrm{UV}} > 2 \times 10^{10}~L_\odot$) and ultra-compact (UV surface brightness $I_{1530\AA} > 10^9~L_\odot~{\mathrm{kpc}}^{-2}$) galaxies (\citealt{2005ApJ...619L..35H}) at redshift $z\sim 0.1-0.3$, includes 6 members with their infrared luminosities similar to local LIRGs and SFRs in the range of $15-110~M_\odot~{\mathrm{yr}^{-1}}$. They have stellar masses of $M_\star \sim 10^{9.8}-10^{10.9}~M_\odot$, and metallicities of $\sim$0.5$-$1 $Z_\odot$ (e.g., see \citealt{2009ApJ...706..203O}). These sources were observed with the \herschel/PACS for \CII\ and with IRAM for \COI.

(7) {\it Intermediate to high redshift galaxies ($z\sim0.3-6$).} This sample of galaxies include 5 (U)LIRGs at redshift $z\sim0.3$ (\citealt{2014ApJ...796...63M}), 12 submillimetre galaxies (SMGs) and QSOs with $z=2-6$ (\citealt{2012MNRAS.425.2203T,2013ARA&A..51..105C,2014ApJ...796...63M,2014ApJ...783...59R,2015Natur.522..455C,2017ApJ...835..110M,2018MNRAS.481...59Z,2019ApJ...871...85L,2020ApJ...895...81R}), and 20 sources ($z\sim1-6$) with the CO observed in the 2$-$1 transition (\citealt{2010ApJ...724..957S,2013ARA&A..51..105C,2014ApJ...780..142F,2014MNRAS.443L..54H,2018ApJ...861...43P,2019ApJ...882..168P,2020MNRAS.494.4090C,2020ApJ...895...81R}). These sources are generally massive systems with $M_\star \gtrsim 10^{10}~M_\odot$ (e.g., \citealt{2011ApJ...740...96H,2012A&A...541A..85M}), and among the 20 objects with $z>4$, eight have $z>5$. As in \cite{2015MNRAS.449.2883G}, we have adopted a CO(2$-$1)-to-\COI\ flux (W~m$^{-2}$) ratio of 7.2 (i.e., a brightness temperature ratio $r_{21}$ of 0.9) to convert the CO(2$-$1) flux to the \COI\ flux. While the adopted $r_{21}$ is a reasonable value based on several observational results (\citealt{2013MNRAS.429.3047B,2014MNRAS.442..558A,2014ApJ...785..149S,2023ApJ...948...44R}), and consistent with that ($r_{21}=0.88\pm0.07$) from turbulence model in \cite{2021ApJ...908...95H}, it might introduce an additional scatter of about 0.14 dex in the CO luminosity given the observed range of $r_{21}\sim0.6-1.2$. Three sources have their oxygen abundances estimated in the literature (HLSW-01: \citealp{2018MNRAS.473...20R}; SDP.81: \citealp{2023A&A...679A.119R}; SPT0418-47: \citealp{2024NatAs...8..368P}) and another three (SWIRE5: \citealp{2019MNRAS.486.5621P}; HDF850.1: \citealp{2023arXiv230904525H}; GN10: \citealp{2024ApJ...961...69S}) have optical lines flux measurements, for which the metallicities are derived and converted to the PP04 calibration.

In total there are 206 galaxies having both \CII\ and CO data, and the typical uncertainties of CO and \CII\ line fluxes are about 20\% and 13\% respectively. Auxiliary data for a part of the sample galaxies (specially for GOALS LIRGs; see \citealt{2017ApJ...846...32D}) are also available, such as the \OIIIFIR, \NIIab\ (\citealt{2016ApJ...819...69Z,2017ApJS..230....1L}), and FIR size (\citealt{2016A&A...591A.136L}). For the CO line luminosity measured in $L_\odot$ (hereafter \LCO), it is related to that expressed in units of \kkmspc\ ($L'_{\mathrm{CO}}$; \citealt{2005ARA&A..43..677S}) as,
\begin{equation}
   L^\prime_{\mathrm{CO}}= 2.04 \times 10^4 L_{\mathrm{CO}}.
\end{equation}

\subsection{Comparison Samples}
To have a more comprehensive understanding of the relation between \CII\ and CO emissions, we also include the following samples of Galactic and extragalactic regions/nuclei and metal-poor, dwarf galaxies for comparison.  

(1) {\it Galactic/Extragalactic regions} (\citealt{1991ApJ...373..423S,2021ApJ...909..204F}). The Galactic data include 9 giant molecular clouds (GMC) and 20 \HII\ regions, and the extragalactic nuclei contain 25 points, consisting of 14 main-sequence (i.e., dust temperature $T_{\mathrm{dust}}<40$~K) and 11 star-forming ($T_{\mathrm{dust}}>40$~K) regions (\citealt{1991ApJ...373..423S}). We further add the 12 main-sequence regions in NGC~7469 presented in \cite{2021ApJ...909..204F}.

(2) {\it Dwarf galaxies} (\citealt{2020A&A...643A.141M,2022AJ....164...44M}). This sample of galaxies include 15 members, thirteen out of which have $Z\sim0.15-0.5~Z_\odot$ and two have $Z\sim0.7~Z_\odot$. For 12 out of these sources, the \CII\ and \COI\ data are provided in \citetalias{2018MNRAS.481.1976Z}. For the remaining 3 galaxies from the Virgo cluster, the CO data are adopted from \cite{2016A&A...590A..27G}, and the \CII\ data are given in \cite{2022AJ....164...44M}. 
 
\section{Results and Discussion}
\subsection{Correlation between \CII\ and \COI\ Line Luminosities}
\label{sect:ciicorelation}
\begin{figure*}[tb]
\centering
\includegraphics[width=0.72\textwidth,bb = 35 7 390 264]{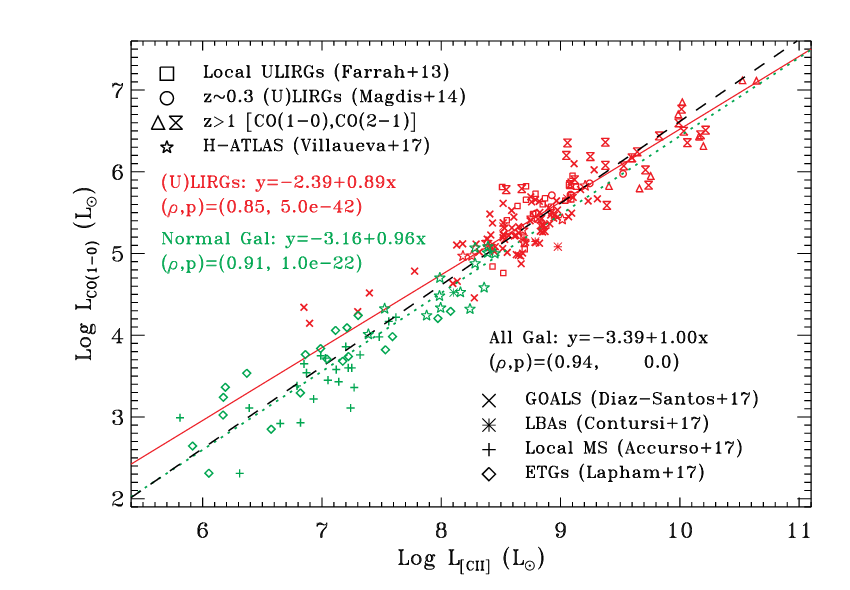}
\caption{\COI\ line luminosity plotted against \CII\ line luminosity. Red and green symbols show (U)LIRGs and normal galaxies, respectively. Open diamonds are early-type galaxies from \cite{2017ApJ...840...51L}, plus signs represent local main-sequence galaxies from \cite{2017MNRAS.470.4750A}, asterisks demonstrate Lyman-break analogs at $z\sim0.2$ from \cite{2017A&A...606A..86C}, crosses and squares respectively display local (U)LIRGs from \cite{2017ApJ...846...32D} and \cite{2013ApJ...776...38F}, stars are the {\it H-ATLAS} star-forming sources at $z\sim0.02-0.2$ from \cite{2015MNRAS.449.2498I}, circles show $z\sim0.3$ (U)LIRGs from \cite{2014ApJ...796...63M}, triangles and bowties are high-redshift ($z>1$) sources, including (U)LIRGs, QSOs and SMGs, respectively observed in \COI\ (\citealt{2013ARA&A..51..105C,2014ApJ...796...63M,2014ApJ...783...59R,2001ApJ...561..766M,2017ApJ...835..110M,2018MNRAS.481...59Z,2019ApJ...871...85L,2020ApJ...895...81R}), and in CO(2$-$1)  (\citealt{2013ARA&A..51..105C,2014ApJ...780..142F,2014MNRAS.443L..54H,2018ApJ...861...43P,2019ApJ...882..168P,2020MNRAS.494.4090C,2020ApJ...895...81R}). The fitted trends are annotated with corresponding correlation coefficients ($\rho$) and probability of null hypothesis ($p$), and are plotted as red solid, green dotted and black dashed lines, for (U)LIRGs, normal galaxies and the whole sample, respectively.}
\label{FigCOvsLCII}
\end{figure*}

In Figure \ref{FigCOvsLCII} we plot the correlation between \LCO\ and \LCII, with red and green symbols respectively representing (U)LIRGs (i.e., $L_{\mathrm{IR}}>10^{11}~L_\odot$) and normal galaxies (i.e., $L_{\mathrm{IR}}<10^{11}~L_\odot$). As shown in the figure, the \COI\ emission is well correlated with the \CII\ line. The corresponding Spearman's rank correlation coefficients, $\rho$, which  assesses how well an arbitrary monotonic function could describe the relationship between two variables, of the trends are 0.85, 0.91, and 0.94 for (U)LIRGs, normal galaxies and the whole sample, respectively, with the probabilities ($p$) for the null hypothesis of no correlation between \LCO\ and \LCII\ are $\approx 0$. 

To investigate the quantitative relationship between \LCO\ and \LCII, we fitted each subsample using an unweighted least-squares bisector with a linear form, as recommended for determining the functional relationship between two parameters by \cite{1990ApJ...364..104I}, i.e.,
\begin{equation}
 \log L_{\mathrm{CO}} (L_\odot) = a+b\log L_{\mathrm{[C\,\scriptsize{\textsc{ii}}]}} (L_\odot),
\end{equation}
The fitted trends are shown as red solid and green dotted lines in Figure \ref{FigCOvsLCII}, respectively for (U)LIRGs and normal galaxies. Using the same method, we also fitted the whole galaxy sample, and the obtained trend is plotted as the black dashed line in Figure \ref{FigCOvsLCII}. 

\begin{deluxetable}{lccccc}[t]
\centering
\tablecaption{Coefficients for Equations (2) and (3)}
\tablewidth{0pt}
\tabletypesize{\scriptsize}
\tablehead{
\colhead{\multirow{2}{*}{Sample}} & \colhead{\multirow{2}{*}{\it{N}}}&\colhead{Intercept}&\colhead{Slope}&\colhead{Scatter}\\
&&($a$)&$(b)$&(dex)
}
\startdata
(U)LIRGs  &148  & $-2.39(0.34)$ &$0.89(0.04)$&0.24\\
Normal   &58 & $-3.16(0.45)$ &$0.96(0.06)$&0.29\\
All &206&$-3.39(0.21)$&$1.00(0.02)$&0.28
\enddata
\tablecomments{Numbers in the parentheses give the $1\sigma$ uncertainties.}
\label{Tabderived}
\end{deluxetable}

\begin{figure}[t]
\centering
\includegraphics[width=0.45\textwidth,bb = 24 39 405 786]{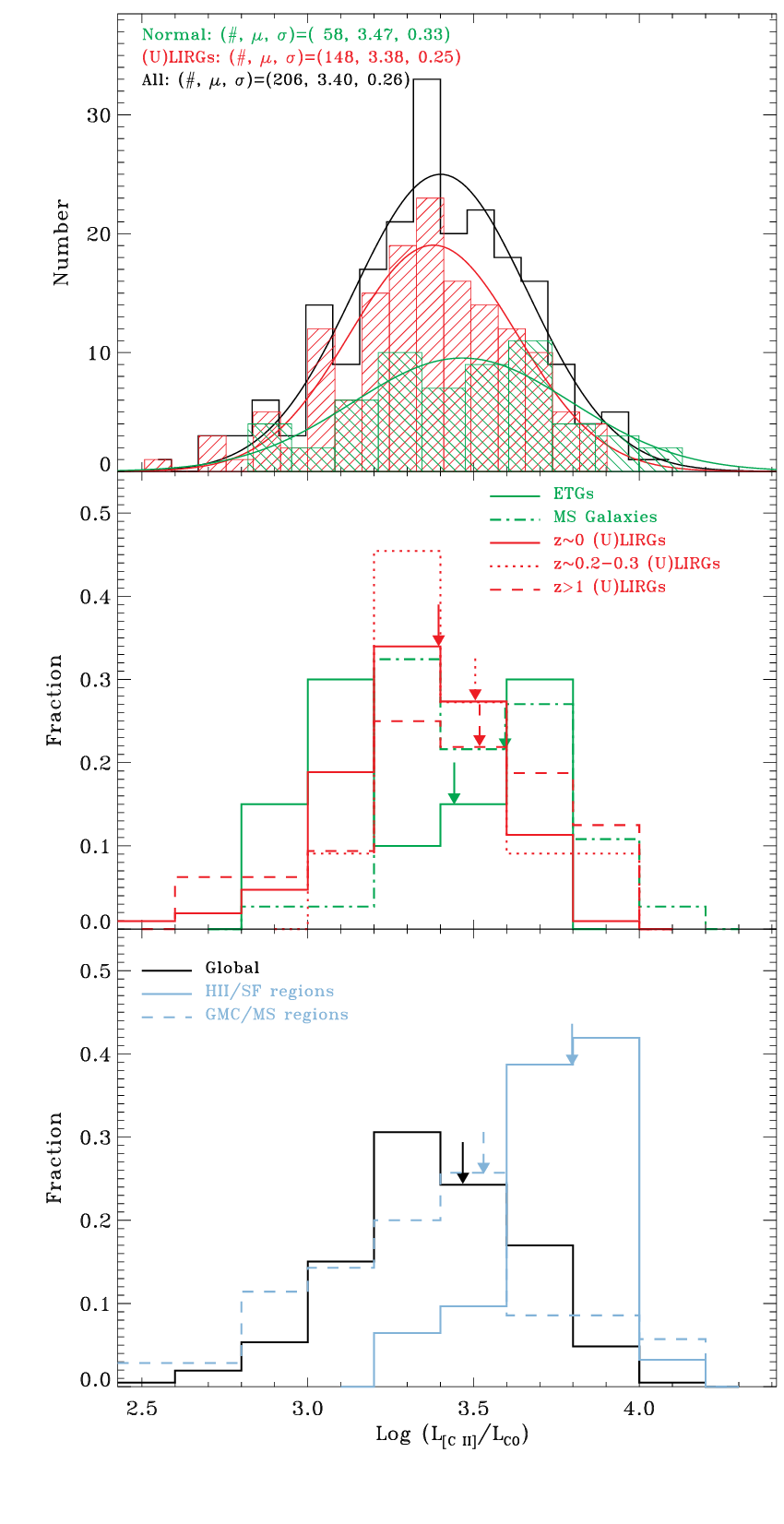}
\caption{Comparison of the distribution of the \CII-to-CO luminosity ratios for various samples. {\it Top}: The whole sample (black), (U)LIRGs (red) and normal galaxies (green), with the result of a Gaussian fit to the data superposed. The number of members, and mean \LCII/\LCO\ ratio and scatter from the Gaussian fit for each sample are also labelled with corresponding colors. {\it Middle}: ETGs (green solid line), MS galaxies (green dashed-dotted line), local (red solid line), $z\sim 0.2-0.3$ (red dotted line) and $z>1$ (red dashed line) (U)LIRGs. {\it Bottom}: Galaxy-integrated results of our whole sample (black solid line), Galactic \HII\ regions and extragalactic starburst nuclei (\citealt{1991ApJ...373..423S}; blue dashed line), and Galactic GMCs and non-starburst nuclei/regions  (\citealt{1991ApJ...373..423S,2021ApJ...909..204F}; blue solid line). The arrows in the middle and bottom panels indicate the corresponding mean values.}
\label{FigCII2CODist}
\end{figure}

Table \ref{Tabderived} lists the number of objects in each sample, the fitting
coefficients, $1\sigma$ errors, and scatters for the \LCII$-$\LCO\ relation. The intercepts shown in the table suggest that the \COI\ emission has a linear correlation with the \CII\ emission. Furthermore, as shown in the middle panel of Figure \ref{FigCII2CODist}, which plots the distributions of the logarithmic \LCII-to-\LCO\ ratios (\LCII/\LCO) for different subsamples with various galaxy types/redshifts, we can see that the largest difference (a factor of $\sim$1.6) of the mean \LCII/\LCO\ (indicated by the corresponding arrows) is between local (U)LIRGs ($2500\pm1300$; red solid line) and MS galaxies ($3900\pm2700$; green dash-dotted line). The (U)LIRG subsamples show a difference of $\lesssim$35 per cent among the three redshift bins, with an increasing \LCII/\LCO\ ratio from $z=0$ to $z>1$ ($3300\pm2200$). We also note that ETGs have a double-peaked distribution, much different from the other subsamples, but the mean \LCII/\LCO\ ratio ($2800\pm1900$) shows relatively small differences ($\lesssim$40\%) compared with the other samples. Nevertheless, we caution that such analysis is possibly limited by the small numbers of the samples, especially for ETGs (20 members) and $z=0.2-0.3$ (U)LIRGs (11 members). 

As mentioned in Section \ref{sec:intro} and already shown in \citet[and references therein]{2020A&A...643A.141M}, however, the situation changes much for dwarf/low-metallicity galaxies, most of which have $L_{\mathrm{[C\,\scriptsize{\textsc{ii}}]}}/L_{\mathrm{CO}} \gtrsim 10^4$ (can be as high as $10^{4.9}$; also see Section \ref{sect:alphaco}), in general much larger than the ratios of massive/metal-rich galaxies. It is also interesting to see that, as demonstrated in the bottom panel of Figure \ref{FigCII2CODist}, the \CII-to-CO ratios on galaxy-wide scale (with a mean value of $2900\pm1900$) are more similar to those of Galactic giant molecular clouds (GMCs; \citealt{1991ApJ...373..423S}) and non-starburst regions like NGC~7469 (\citealt{2021ApJ...909..204F}), for which the mean ratio ($3400\pm3500$) is $\sim$2 times smaller than that ($6300\pm2400$) of the Galactic \HII\ regions and starburst nuclei presented in \citet{1991ApJ...373..423S}.

Our results indicate that, on galactic scale, massive/metal-rich galaxies under different physical conditions (from quiescent ETGs to extreme SF ULIRGs/SMGs) and at different redshifts (from $z=0$ to $z\sim6$) seem to share almost identical \LCO$-$\LCII\ relation, with comparable scatters of a factor $\la 2$. The rms deviations of the data from the fitted relationships in \LCO\ are 0.24, 0.29 and 0.28 dex for (U)LIRGs, normal galaxies and the whole sample, respectively. We also note that the sources with CO(2$-$1)-converted \LCO\ have a larger scatter of 0.37 dex, confirming the additional uncertainty introduced by $r_{21}$ as mentioned in Section \ref{sec:sample}.

 Therefore, the \COI\ and \CII\ emissions should have a strong physical connection as illustrated by both observational (e.g., \citealt{2014A&A...572A..45V,2017ApJ...839..107P}) and theoretical (\citealt{2017MNRAS.464.3315A,2020A&A...643A.141M,2021MNRAS.502.2701B}) works, even though the latter has additional origins other than PDRs. Given that the \COI\ emission is a commonly used tracer of \MHH, this scaling, established without any assumptions about the physical state of the gas, further confirms that the \CII\ line is able to trace the total molecular gas for galaxies at local and high-$z$ Universe. This is a powerful and particularly useful tool for probing the total molecular gas content in galaxies at $z>4$ (e.g., \citetalias{2018MNRAS.481.1976Z}; \citealt{2020A&A...643A...5D}), because that other tracers of molecular gas become increasing difficult to observe with ground-based facilities, whereas the \CII\ line is redshifted into the millimeter atmospheric windows accessible to ground (sub-)mm telescopes.

To convert the integrated luminosity of the \CII\ line to the total molecular gas mass, we can utilize the \LCO-\LCII\ relations given by Equation (2), in combination with the CO-to-\HH\ conversion factor, $\alpha_{\mathrm{CO}} [M_\odot~({\mathrm{K~km~s^{-1}~pc^2}})^{-1}] \equiv M_{\mathrm{mol}}/L^\prime_{\mathrm{CO}}$ (including a factor 1.36 to account for helium). Substituting \LCO\ in Equation (2) with Equation (1), we obtain the following \LCII-\MHH\ relation,  
\begin{equation}
     \log M_{\mathrm{mol}} (M_\odot) =\log (2.04\times10^4\alpha_{\mathrm{CO}}) + a + b \log L_{\mathrm{[C\,\scriptsize{\textsc{ii}}]}} (L_\odot).
\end{equation}

Given the strong linear relation between \LCO\ and \LCII, alternatively, it is natural and convenient to define a \CII-to-\HH\ conversion factor by analogy with $\alpha_{\mathrm{CO}}$ as in \citetalias{2018MNRAS.481.1976Z}, namely,
\begin{equation}
    \alpha_{[{\mathrm{C}}\,\scriptsize{\textsc{ii}}]} (M_\odot/L_\odot) \equiv  M_{\mathrm{mol}}/L_{[{\mathrm{C}}\,\scriptsize{\textsc{ii}}]}.
\end{equation}
By substituting Equation (1) and $\alpha_{\mathrm{CO}}$ for \MHH, Equation (4) can be rewritten as, 
\begin{equation}
    \alpha_{[{\mathrm{C}}\,\scriptsize{\textsc{ii}}]} =  2.04 \times10^4 \alpha_{\mathrm{CO}}/(L_{[{\mathrm{C}}\,\scriptsize{\textsc{ii}}]}/L_{\mathrm{CO}}).
\end{equation}

To assess the value of \alphaCII, we plot the distributions of the logarithmic \LCII-to-\LCO\ ratios (\LCII/\LCO) in the top panel of Figure \ref{FigCII2CODist} for the whole sample and the two subsamples, with the results of a Gaussian fit to the data superposed. The number of members, and the mean value and scatter from the Gaussian fit are also labelled. In general a normal distribution can well describe the observed \LCII/\LCO\ ratios. According to the results of the Gaussian fit shown in the top panel of Figure \ref{FigCII2CODist}, $\alpha_{[{\mathrm{C}}\,\scriptsize{\textsc{ii}}]}$ is related with $\alpha_{\mathrm{CO}}$ via, 
\begin{equation} 
    \alpha_{[{\mathrm{C}}\,\scriptsize{\textsc{ii}}]} = [8.50, 6.91, 8.12]\times \alpha_{\mathrm{CO}}. 
\end{equation}
for {\it (U)LIRGs}, {\it normal galaxies} and the {\it whole sample}, respectively. The corresponding uncertainties of $\alpha_{[{\mathrm{C}}\,\scriptsize{\textsc{ii}}]}$, caused by the scatter of \LCII/\LCO, are 0.25 dex, 0.33 dex and 0.26 dex.  

\subsection{The choice of the CO-to-\HH\ conversion factor}
\label{sect:alphaco}

\begin{deluxetable}{lccc}[t]
\centering
\tablecaption{CO- and \CII-to-\HH\ Conversion Factors}
\tablewidth{0pt}
\tabletypesize{\scriptsize}
\tablehead{
\colhead{\multirow{2}{*}{Sample}} &\colhead{$\alpha_{\mathrm{CO}}$}&\colhead{\alphaCII}&\colhead{Uncertainty}\\
&($M_\odot~(\mathrm{K~km~s^{-1}~pc^2})^{-1}$)&($M_\odot/L_\odot$)&(dex)
}
\startdata
(U)LIRGs   & 4.0 &34.0&0.25\\
Normal   & 4.7 &32.5&0.33\\
All &4.3&34.9&0.26
\enddata
\tablecomments{(1) $\alpha_{\mathrm{CO}}$ is adopted from \citet{2022MNRAS.517..962D}. (2) The uncertainty of \alphaCII\ is estimated from the scatter of the \LCII/\LCO\ luminosity ratios.}
\label{Tableconverionfactor}
\end{deluxetable}

As shown in Equations (3) and (6), the conversion of \LCII\ to \MHH\ relies entirely on the CO-to-\HH\ conversion factor $\alpha_{\mathrm{CO}}$, in consequence of utilizing the CO emission to calibrate \MHH. However, the conversion factor of CO is a long and debated discussion in the literature. It can vary by a factor of 3$-$5, from 0.8$-$1.5 $M_\odot~({\mathrm{K~km~s^{-1}~pc^2}})^{-1}$ for extreme local and high-$z$ starbursts like ULIRGs/SMGs (\citealt{1998ApJ...507..615D}) to 4.36 $M_\odot~({\mathrm{K~km~s^{-1}~pc^2}})^{-1}$ for Milky Way-like galaxies (e.g., see \citealt{2013ARA&A..51..207B} for a review). 

In recent years, however, an increasing number of studies, based on dust determinations and other molecular gas tracer such as \CI, have reported a lack of such bi-modality in $\alpha_{\mathrm{CO}}$ (e.g., \citealt{2012ApJ...760....6M,2014MNRAS.441.1017R,2015ApJ...800...20G,2022MNRAS.517..962D}). \cite{2022MNRAS.517..962D} have estimated $\alpha_{\mathrm{CO}}$ through a self-consistent cross-calibration between three different molecular gas tracers (i.e., CO, \CI\ and sub-millimetre dust continuum) for a large, heterogeneous sample of galaxies up to $z\sim6$. They find that the difference of $\alpha_{\mathrm{CO}}$ between IR-luminous (i.e., $L_{\mathrm{IR}}>10^{11}~L_\odot$) and less luminous ($L_{\mathrm{IR}}<10^{11}~L_\odot$) galaxies is very small – around 10$-$20 per cent in the mean, rather than the often assumed factor of $\sim$3$–$5 as mentioned above. 
Our sample of galaxies, comprising different types of objects, demonstrates that they very likely share a universal \LCII-\LCO\ relation. Therefore, a bimodal $\alpha_{\mathrm{CO}}$ would also lead to a bimodal \CII\ conversion factor. However, a bimodal $\alpha_{[{\mathrm{C}}\,\scriptsize{\textsc{ii}}]}$ seems not supported by any observational evidence (e.g., \citetalias{2018MNRAS.481.1976Z}; \citealt{2020A&A...643A...5D}) or by simulations (e.g., \citealt{2020A&A...643A.141M,2022ApJ...929...92V}) so far, and thus is unlikely applicable to our sample galaxies.

In light of the above considerations and given the appreciable overlap between our and Dunne et al. 2022's samples, we adopt the conversion factor $\alpha_{\mathrm{CO}}$ of 4.0 and 4.7~$M_\odot~(\mathrm{K~km~s^{-1}~pc^2})^{-1}$ (including the contribution of helium; their Table 7), respectively for (U)LIRGs and normal galaxies. For the whole sample, the mean value of 4.3~$M_\odot~(\mathrm{K~km~s^{-1}~pc^2})^{-1}$ is used. Therefore, the corresponding $\alpha_{[{\mathrm{C}}\,\scriptsize{\textsc{ii}}]}$ are 34.0, 32.5 and 34.9~$M_\odot/L_\odot$ , respectively for (U)LIRGs, normal galaxies and the whole sample (see Table \ref{Tableconverionfactor}). For ETGs only, we obtain $\alpha_{[{\mathrm{C}}\,\scriptsize{\textsc{ii}}],\,{\mathrm{ETG}}}$ of 34.2~$M_\odot/L_\odot$ with an uncertainty of 0.3 dex using the mean \CII-to-CO ratio of $2800\pm1900$ given above and $\alpha_{\mathrm{CO}}$ of 4.7~$M_\odot~(\mathrm{K~km~s^{-1}~pc^2})^{-1}$, agreeing well with the other samples. Given the small differences ($<$10 per cent) between these samples, we conclude that $\alpha_{[{\mathrm{C}}\,\scriptsize{\textsc{ii}}]}$ is very likely universal, irrespective of galaxy types (e.g., intensity of star-forming activities, as already suggested by \citetalias{2018MNRAS.481.1976Z}) and redshifts.

Our derived $\alpha_{[{\mathrm{C}}\,\scriptsize{\textsc{ii}}]}$ are very similar to that (31~$M_\odot/L_\odot$) in \citetalias{2018MNRAS.481.1976Z}, though they have used a different approach to obtain the molecular gas mass (i.e., metallicity-dependent $\alpha_{\mathrm{CO}}$ and KS relation). This is not unreasonable because the sample galaxies in \citetalias{2018MNRAS.481.1976Z} are dominated by main-sequence (MS) objects. For {\it starbursts}, however, \citetalias{2018MNRAS.481.1976Z} obtained $\alpha_{[{\mathrm{C}}\,\scriptsize{\textsc{ii}}]}=22~M_\odot/L_\odot$ using a sample consisting of about 10 galaxies. Their value is $\sim$35\% smaller than our result for (U)LIRGs. This difference might be caused by the samples (our sample is larger by one order of magnitude; see Figure 8 in \citetalias{2018MNRAS.481.1976Z}) and CO conversion factors used between the two works (for these sources, \citetalias{2018MNRAS.481.1976Z} adopted $\alpha_{\mathrm{CO}}$ that has been derived for each source based their dynamical mass from \citealt{2016MNRAS.457.4406A}).

\begin{figure}[t]
\centering
\includegraphics[width=0.45\textwidth,bb = 24 9 549 435]{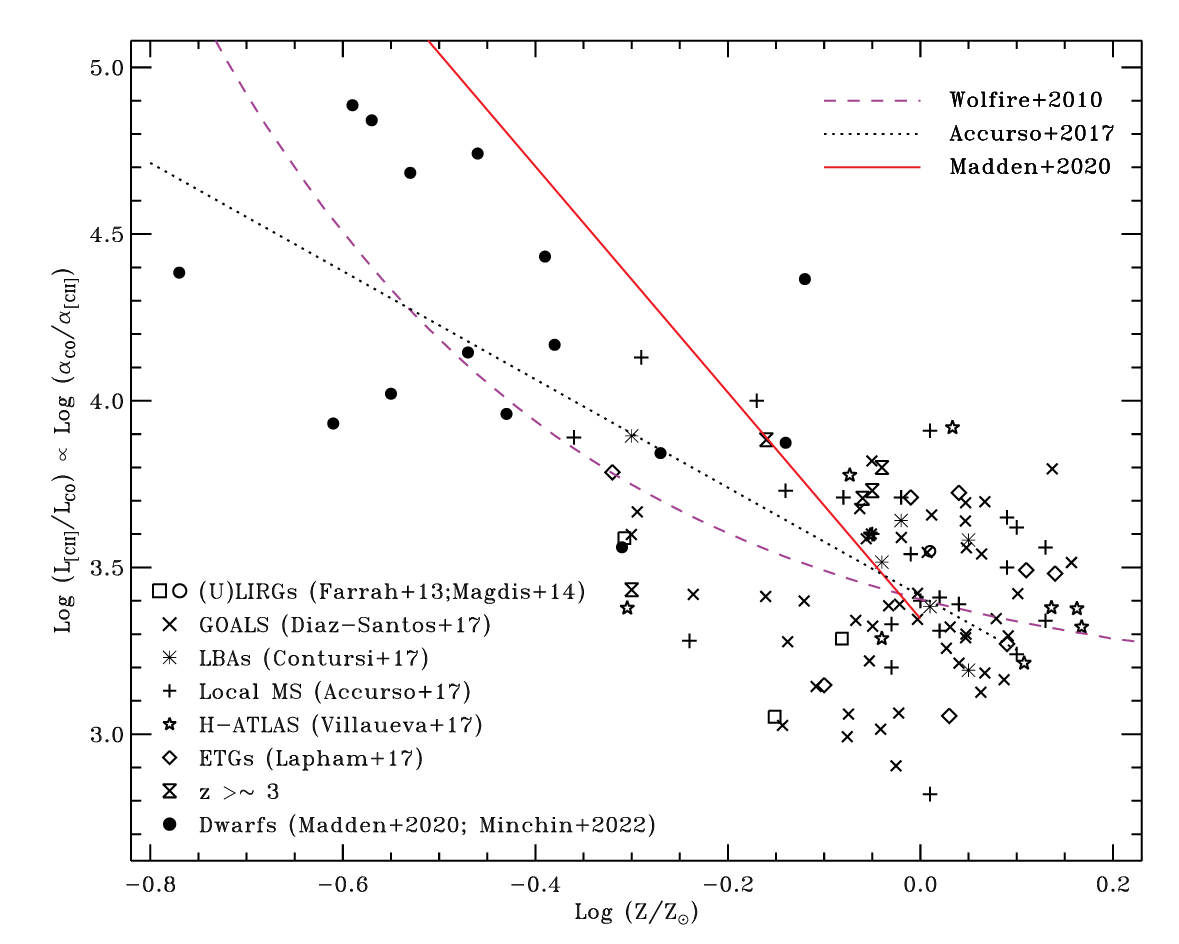}
\caption{\CII-to-CO luminosity ratio plotted versus metallicity for dwarf (\citealt{2015A&A...578A..53C,2016A&A...590A..27G,2020A&A...643A.141M,2022AJ....164...44M}) and massive galaxies. QSOs and AGNs are excluded from the plot. For comparison, the overplotted lines show $\alpha_{\mathrm{CO}}$ as a function of metallicity from \citet[purple dashed line]{2010ApJ...716.1191W}, \citet[black dotted line]{2017MNRAS.470.4750A}, and \citet[red solid line]{2020A&A...643A.141M}, assuming a constant \alphaCII\ of 34.9~$M_\odot/L_\odot$.}
\label{FigCII2CO_metallicity}
\end{figure}

Since \citetalias{2018MNRAS.481.1976Z} sample includes dwarf galaxies with low  metallicities, these results suggest that $\alpha_{[{\mathrm{C}}\,\scriptsize{\textsc{ii}}]}$ seems to be independent on metal abundance over the metallicity range ($12 +\log (\mathrm{O/H})=7.9-8.9$) explored, as discussed in \citetalias{2018MNRAS.481.1976Z}. With a larger sample in hand, here we further investigate whether a metallicity-dependent $\alpha_{\mathrm{CO}}$ and a constant $\alpha_{[{\mathrm{C}}\,\scriptsize{\textsc{ii}}]}$ can reproduce the observed \LCII/\LCO\ ratios. To this end, we plots \LCII/\LCO, which is equal to $\alpha_{\mathrm{CO}}/\alpha_{[{\mathrm{C}}\,\scriptsize{\textsc{ii}}]}$ according to the definitions, as a function of $Z$, for the dwarf (\citealt{2015A&A...578A..53C,2016A&A...590A..27G,2020A&A...643A.141M,2022AJ....164...44M}) and our massive galaxies with metallicity measurements in Figure~\ref{FigCII2CO_metallicity}. For comparison, we also show the $\alpha_{\mathrm{CO}}$-$Z$ relations from \citet[purple dashed line]{2010ApJ...716.1191W}, \citet[black dotted line]{2017MNRAS.470.4750A}, and \citet[red solid line]{2020A&A...643A.141M}, assuming a constant \alphaCII\ of 34.9~$M_\odot/L_\odot$. 

As we can see from the figure, indeed, both the \citet{2010ApJ...716.1191W} and \citet{2017MNRAS.470.4750A} relations, in combination with a constant \alphaCII, follow the observed \LCII/\LCO-$Z$ trend. However, the \citet{2020A&A...643A.141M} relation starts to over-predict the \CII-to-CO ratio at $Z\sim0.6Z_\odot$, which might be caused by their larger  $\alpha_{\mathrm{CO}}$ at lower metallicity. We can further check this by calculating mean \alphaCII, which are $44\pm33$ and $58\pm69$~$M_\odot/L_\odot$, using the $Z$-dependent $\alpha_{\mathrm{CO}}$ from \cite{2017MNRAS.470.4750A} and \cite{2020A&A...643A.141M}, respectively, for our subsample with measured metallicities (excluding literature dwarf data; otherwise, the derived $\alpha_{\mathrm{CO}}=43\pm32$ and $87\pm169$~$M_\odot/L_\odot$, respectively). Therefore, it is difficult to draw a solid conclusion whether \alphaCII\ is insensitive to metallicity based on the current method and data set, due to the large uncertainty of $\alpha_{\mathrm{CO}}$ for low-$Z$ sources as already mentioned in \S2.1.

\subsection{Origin of the \LCII-\LCO\ relation}
\label{sect:origin} 
Given the multiple phases (ionized, neutral atomic and molecular) and powering sources (i.e., far-UV photons, cosmic rays, X-rays, turbulence and shocks; e.g. see a review in \citealt{2022ARA&A..60..247W}) of the \CII\ line, it is remarkable that there exists a linear correlation between the \COI\ and \CII\ emissions. Therefore, it is intriguing to explore the reason(s) resulting in such a relation.

Using numerical simulations for star-forming, main-sequence galaxies at $z\sim6$, \cite{2022ApJ...929...92V} ascribed this correlation (\MHH-\LCII; but sublinear in their work) to the KS law (\citealt{1998ARA&A..36..189K}; also see \citealt{2015ApJ...805...31L} for a sample including 2$\times$ more (U)LIRGs), a tight relationship between SFR and molecular gas mass. In fact, the powering source of the \CII\ emission in star-forming galaxies should be dominated by the current SF activity (e.g., far-UV photons emitted by massive stars, SF-related cosmic rays and shocks; see, e.g., \citealt{1985ApJ...289..803S,1991ApJ...373..423S,2011A&A...532A.152M}), which is, in turn, correlated with the gas content of a galaxy.

\begin{figure*}[t]
\centering
\includegraphics[width=0.98\textwidth,bb = 27 28 674 256]{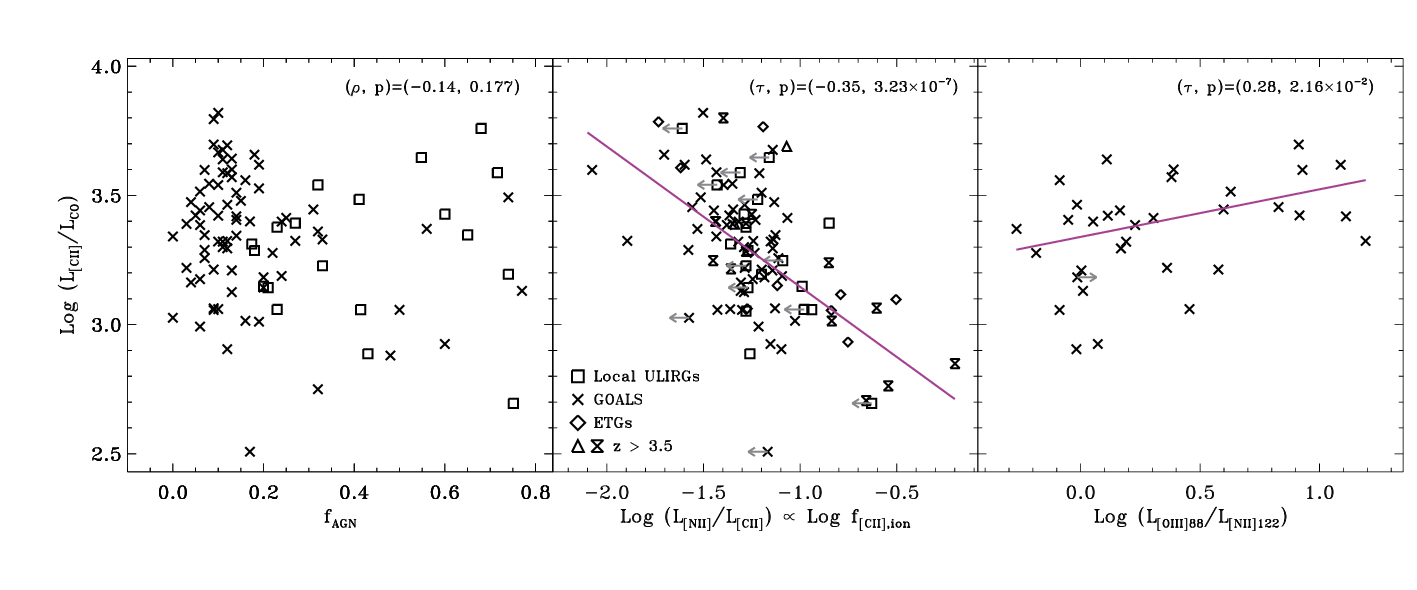}
\caption{\LCII/\LCO\ plotted vs. $f_{\mathrm{AGN}}$ ({\it left}), \LNII/\LCII\ ({\it middel}), and \OIIIFIR-to-\NIIbc\ flux ratio. Arrows in the middle and right panels respectively indicate upper and lower limits of the $x$-axis. The solid lines are linear fits to the data. The correlation coefficient and probability of null hypothesis are also given in each panel.}
\label{FigCII2CO_fagn}
\end{figure*}

However, our sample is a mixture of objects having (1) different levels of AGN activity; and (2) vastly varied SFEs: from low-SFE ETGs to extremely high-SFE ULIRGs. These gas-rich ETGs deviate systematically from the KS relation (\citealt{2014MNRAS.444.3427D}), whereas ULIRGs have shown a well-known ``\CII/FIR" deficit, namely, their \CII-to-FIR continuum luminosity ratios (\LCII/\LFIR) are smaller by as much as an order of magnitude compared to local less luminous galaxies (e.g., \citealt{2001ApJ...561..766M,2003ApJ...594..758L,2015MNRAS.449.2883G,2016ApJ...824..146Z,2017ApJ...846...32D}). Furthermore, the \CII\ emission in some ETGs is dominated by diffuse ionized gas (\citealt{2017ApJ...840...51L}; Liu et al., in preparation), which might not relate to the current star-forming activity but is ionized by emission from old, hot stars such as postasymptotic giant branch stars (e.g. \citealt{1994A&A...292...13B}). Therefore, it raises the question about whether the explanation in \cite{2022ApJ...929...92V} is also applicable to our results.

\subsubsection{Where Does the \CII\ Emission Come From?}
\label{sect:cii_origin}

To further investigate the origin of the \LCII-\LCO\ relation, first we explore the gas where the \CII\ emission mainly arising from. To this end, firstly we plot the \LCII/\LCO\ ratio as a function of the potential contribution of an AGN to the bolometric luminosity ($f_{\mathrm{AGN}}$; see \citealt{2017ApJ...846...32D} for the calculation), as shown in the left panel of Figure \ref{FigCII2CO_fagn}, for the GOALS LIRGs and local ULIRGs. There is no correlation ($\rho=-0.14$ and $p=0.177$) between these two parameters, suggesting that the \CII\ emission in these sources is not dominated by the X-ray dissociation regions (XDRs). This argument is further evidenced by the fact that the \OIFIR-to-\CII\ (non-ionized gas) luminosity ratios for almost all of these (U)LIRGs are less than unity (\citealt{2013ApJ...776...38F,2017ApJ...846...32D}), about an order of magnitude smaller than the values ($\gtrsim$10; \citealt{2007A&A...461..793M}) predicted by the XDR models with an incident X-ray flux of $F_X(1-100\,\mathrm{kev})\approx 10$~erg~s$^{-1}$~cm$^{-2}$ (i.e., illuminated by a central source with the X-ray luminosity of 10$^{43}$~erg~s$^{-1}$ at a distance of 100~pc without extinction) and gas density of $n=10^3-10^{6.5}$~cm$^{-3}$.

Secondly, we show \LCII/\LCO\ versus \LNIIab/\LCII, the observed \NIIab-to-\CII\ luminosity ratio, which is a proxy of the \CII\ fractional contribution of the ionized gas ($f_{\mathrm{[C\,\scriptsize{\textsc{ii}}],ion}}$; e.g., \citealt{2006ApJ...652L.125O}), in the middle panel of Figure \ref{FigCII2CO_fagn} for those sources with available \NIIab\ measurements (\citealt{2014A&A...565A..59D,2014ApJ...783...59R,2016ApJ...832..151P,2018ApJ...861...43P,2019ApJ...882..168P,2016ApJ...819...69Z,2020MNRAS.494.4090C,2022ApJ...928..179L}). About ninety percent of these galaxies have their \LNIIab/\LCII\ smaller than $\sim$0.1, suggesting $f_{\mathrm{[C\,\scriptsize{\textsc{ii}}],ion}} \lesssim 0.3$ using the carbon and nitrogen abundances in diffuse gas of the MilkyWay (\citealt{1996ARA&A..34..279S}). Those $\sim$10 galaxies having $L_{\mathrm{[N\,\scriptsize{\textsc{ii}}]205\,\mu m}}/L_{\mathrm{[C\,\scriptsize{\textsc{ii}}]}}>0.1$, however, do not show higher- but lower-than-average \LCII/\LCO\ ratios, and we will discuss this in more detail in \S\ref{sec:ionizedcontribution}. Our result indicates that the \CII\ emission in the vast majority of our sample sources should not be dominated by the ionized gas. 

Thirdly, we display \LCII/\LCO\ versus the \OIIIFIR-to-\NIIbc\ luminosity ratio (\LOIII/\LNIIbc) in the right panel of Figure \ref{FigCII2CO_fagn}, for the 33 GOALS LIRGs that have measured fluxes for both lines (\citealt{2017ApJ...846...32D}). Since the ionization potentials of the \NIIbc\ and \OIIIFIR\ lines differ by more than a factor of two from 14.5~eV to 35~eV, and they have similar critical densities, the line ratio \LOIII/\LNIIbc\ can be a good indicator of the hardness of the radiation field (e.g., \citealt{2011ApJ...740L..29F,2018MNRAS.481...59Z}) based on the \cite{1985ApJS...57..349R} H~{\textsc{ii}} models. As we can see from the figure, the Kendall’s $\tau$ correlation coefficient, computed using the {\it cenken} function in the {\it NADA} package within the public domain {\it $\mathbf{R}$} statistical software environment{\footnote{\url{http://www.R-project.org/}}}, is $\tau = 0.28$, with a $p$-value of $2.2\times10^{-2}$, indicating that there only exists a very weak dependence (slope of 0.18) of \LCII/\LCO\ on \LOIII/\LNIIbc. This result further suggests that the \CII\ emission should not mainly come from the ionized gas.

In summary, PDRs dominate the \CII\ emission for the vast majority of our sample galaxies. However, the \LCII-SFR relation has a much larger scatter of $0.4-0.6$ dex, both for simulated (e.g., \citealt{2015ApJ...813...36V,2022ApJ...929...92V}) and observational data sets (e.g, \citealt{2014A&A...568A..62D}; \citetalias{2018MNRAS.481.1976Z}; \citealt{2022ApJ...929...92V}), than that ($\lesssim$0.3 dex) of the \LCII-\MHH\ relation, indicating that \LCII\ traces \MHH\ better than SFR (\citetalias{2018MNRAS.481.1976Z}). Therefore, it is very likely that the correlation between \CII\ and CO emission is mainly caused by their co-spatial origins. However, it is still needed to answer why the relationship is almost linear and so tight over about five order of magnitude in \COI\ (\CII) luminosity despite different physical conditions.

\subsubsection{Comparison to PDR Models}
\label{sect:pdr_model}


\begin{figure}[t]
\centering
\includegraphics[width=0.45\textwidth,bb = 19 7 390 264]{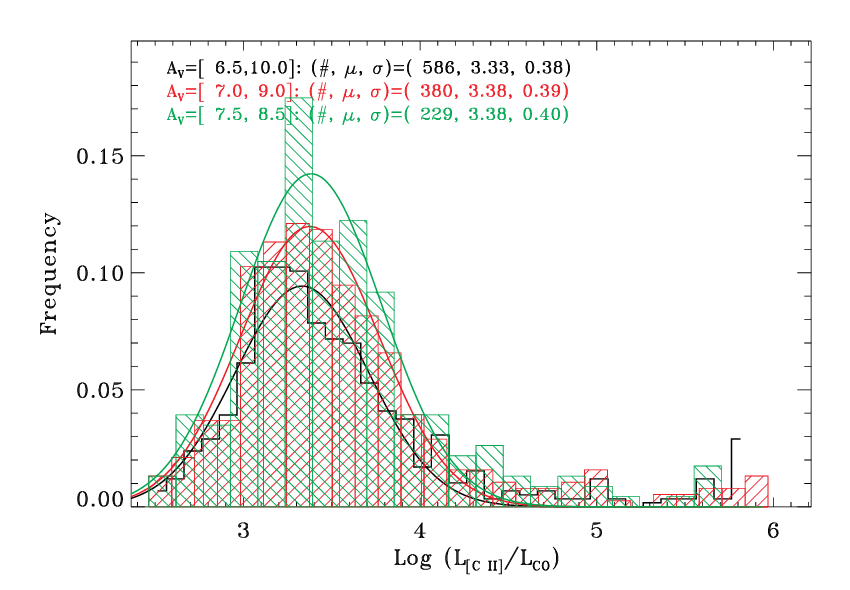}
\caption{Distribution of \LCII/\LCO\ from \cite{2020A&A...643A.141M} PDR modelsets ($Z=0.05-1~Z_\odot$, $G_0=10^{1.25}-10^{4.06}$, and $n=10^1-10^4$~cm$^{-3}$), overlaid with the Gaussian fitting results. The black, red and green colors show the results of models having $A_V=6.5-10$, $7-9$ and $7.5-8.5$ (all with mean $A_V\sim8$), respectively.}
\label{FigPDRmodel_Madden}
\end{figure}


\begin{figure*}[t]
\centering
\includegraphics[width=0.95\textwidth,bb = 51 302 483 462]{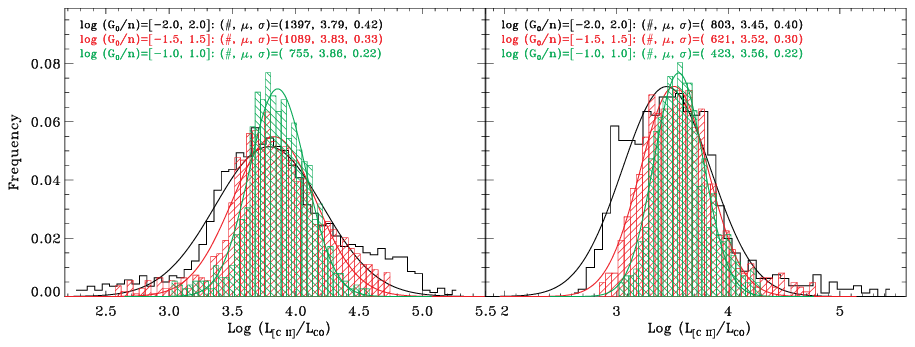}
\caption{Distributions of \LCII/\LCO\ from the ``WK2020" ({\it left}; \citealt{2023AJ....165...25P}) and ``KOSMA-tau" ({\it right}; \citealt{2013A&A...549A..85R,2022A&A...664A..67R}) PDR modelsets ($Z=Z_\odot$ and $\zeta_{\mathrm{CR}} = 2\times10^{-16}~\mathrm{s}^{-1}$), overlaid with the Gaussian fitting results. The black, red and green colors show the results of models having $G_0/n=10^{-2}-10^{2}$, $10^{-1.5}-10^{1.5}$ and $10^{-1}-10^{1}$~cm$^3$, respectively.}
\label{FigPDRmodel}
\end{figure*}

In this section, we compare the observed line ratios with standard PDR models published in the literature. As it has been discussed in detail in \cite{1999ApJ...527..795K}, \cite{2010ApJ...716.1191W} and \cite{2020A&A...643A.141M}, the \LCII/\LCO\ ratio of a molecular cloud depends on gas density $n$, incident FUV radiation field $G_0$, metallicity $Z$ and column density $N$ (and thus visual extinction $A_V$ as $A_V\propto ZN$ assuming a linear relation between dust-to-gas ratio and metallicity; e.g., \citealt{1999ApJ...527..795K,2010ApJ...716.1191W}). Furthermore, \LCII/\LCO\ is most sensitive to $A_V$ (\citealt{2010ApJ...716.1191W,2020A&A...643A.141M}).

Figure \ref{FigPDRmodel_Madden} plots the distributions of \LCII/\LCO\ adopted from the \cite{2020A&A...643A.141M} PDR modelset for different $A_V$ ranges. For a given mean $A_V$ of $\sim$8~mag, which is the typical value of the Galaxy (e.g., \citealt{2010ApJ...716.1191W}), the modelled \LCII/\LCO\ ratios show a log-normal distribution with a nearly constant mean value ($\mu$) very close to our observed one, indicating that our sample galaxies may have an average $A_V$ similar to the Galaxy. From the figure we can also see that the dispersion ($\sigma$) does not change with the range of $A_V$, and is a bit larger than the observed value, suggesting that it needs some other parameters to be responsible for the scatter of the observed \LCII/\LCO\ distribution.

The ratio of $G_0$ to $n$, $G_0/n$, an important parameter in the chemistry and physics of PDRs (\citealt{1985ApJ...291..722T}), to a great extent determines the \LCII/\LCO\ ratio (e.g., \citealt{1998ApJ...498..735P,1999ApJ...527..795K,2007A&A...461..793M,2013A&A...549A..85R}) for fixed $Z$ and fixed $N$ (and thus fixed $A_V$). To explore whether the observed scatter can be attributed to the variation of $G_0/n$, we also compare the observed \LCII/\LCO\ to the PDR models computed with $Z=Z_\odot$ and much larger parameter space in $G_0$ and $n$ in Figure \ref{FigPDRmodel}. The left and right panels display the distributions of \LCII/\LCO\ for three different $G_0/n$ ranges, adopted from the Wolfire$-$Kaufman ``WK2020" (\citealt{2023AJ....165...25P}) and KOSMA-tau ``kt2013wd01-7" clumpy (\citealt{2013A&A...549A..85R,2022A&A...664A..67R}) modelsets\footnote{PDRT Toolbox website: \url{https://dustem.astro.umd.edu}}, respectively. The ``WK2020" modelset includes model grids computed with $G_0=10^{-0.5}-10^{6.5}$ (in units of the Habing field of $1.6\times10^{-3}$~erg~s$^{-1}$~cm$^{-2}$; \citealt{1968BAN....19..421H}), $n=10^1-10^7$~cm$^{-3}$, $Z=Z_\odot$, $A_V=7$~mag, and a cosmic ray ionization rate of $\zeta_{\mathrm{CR}} = 2\times10^{-16}~\mathrm{s}^{-1}$ (\citealt{2023AJ....165...25P}) by assuming a plane-parallel geometry, and the ``kt2013wd01-7" clumpy modelset contains models computed with $G_0\sim10^{0.2}-10^{6.2}$, $n=10^3-10^7$~cm$^{-3}$, $Z=Z_\odot$, $\zeta_{\mathrm{CR}} = 2\times10^{-16}~\mathrm{s}^{-1}$, and maximum clump mass $=1000~M_\odot$ (which determines $A_V$) by using an ensemble of spherical clumps. 

Generally, the modelled \LCII/\LCO\ ratios also show a lognormal-like distribution. It can be seen from Figure \ref{FigPDRmodel} that the dispersion $\sigma$ varies with the range of $G_0/n$ considered. It increases from 0.2 dex to 0.4 dex as the $G_0/n$ range changes from $10^{-1.0}-10^{1.0}$ to $10^{-2.0}-10^{2.0}$~cm$^3$ for both modelsets. Therefore, the gas conditions in the PDRs of different galaxies (with similar metallicities) can be largely varied, but variations in the resulted \LCII/\LCO\ ratio might be as small as 0.2$-$0.4 dex.

Compared to our sample presented in \S\ref{sect:ciicorelation}, we find that $\sigma$ ($\sim$0.3 dex) from $G_0/n=10^{-1.5}-10^{1.5}$~cm$^3$ (red lines in Figure \ref{FigPDRmodel}) is well consistent with the observed value. This is understandable since the derived $G_0/n$ is usually in this range for local and high-$z$ star-forming galaxies and AGNs/QSOs (e.g., \citealt{2001ApJ...561..766M,2016ApJ...824..146Z,2017A&A...606A..86C,2017ApJ...846...32D,2017ApJ...837...12W,2019ApJ...875....3L,2020ApJ...889L..11R,2021A&A...652A..66P,2022A&A...662A..60D}). These results suggest that the tight and linear \LCII-\LCO\ relation could be a direct consequence of the effect that PDRs dominate the origin  (and therefore the close physical association between the \CII\ and CO emitting gas) of the \CII\ emission.

\subsection{Scatter of the \LCII/\LCO\ Ratio}
\label{sect:systematic}
\begin{figure}[t]
\centering
\includegraphics[width=0.45\textwidth,bb = 38 34 358 674]{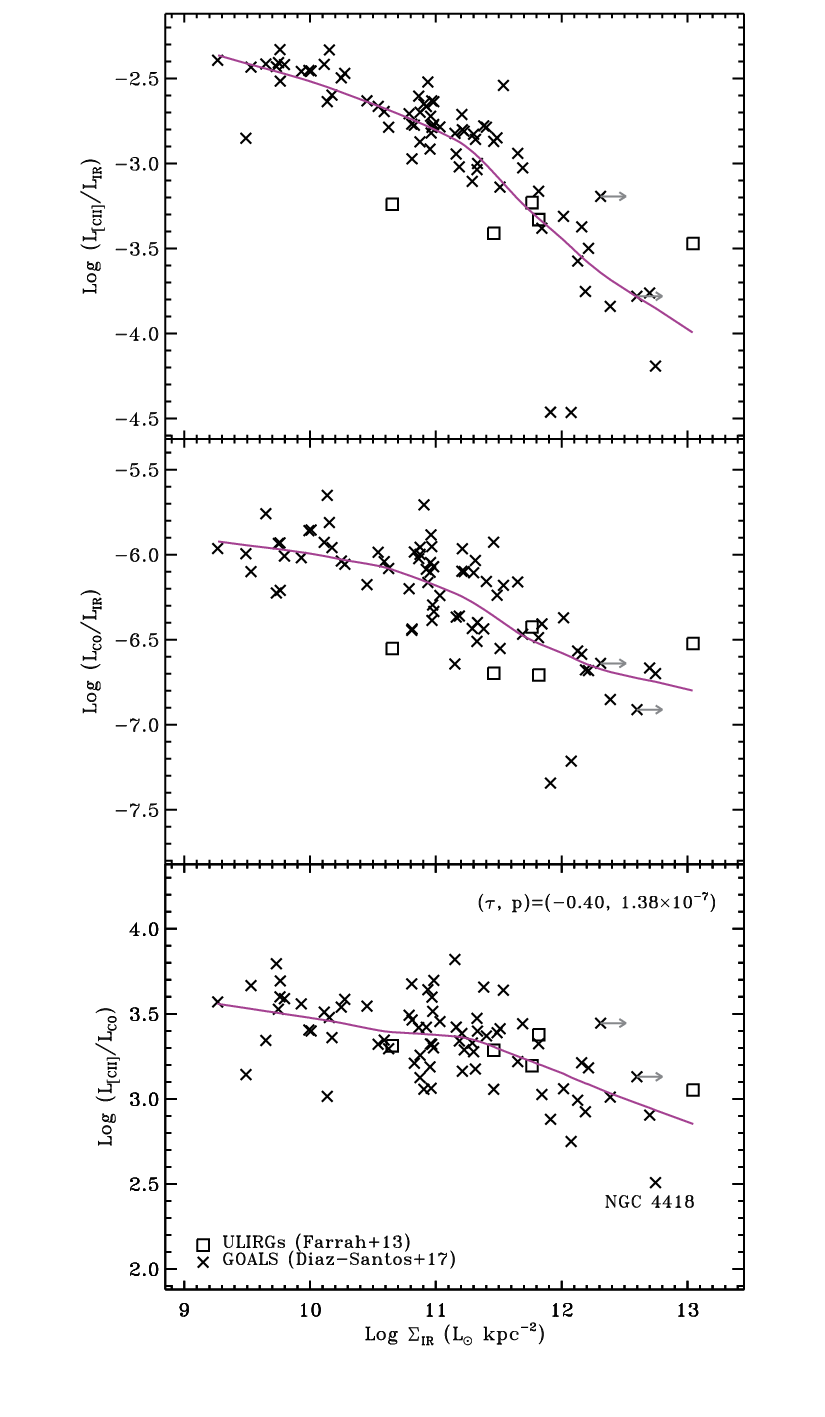}
\caption{Infrared luminosity surface density $\Sigma_{\mathrm{IR}}$ plotted vs. \LCII/\LIR\ ({\it top}), \LCO/\LIR\ ({\it middle}) and \LCII/\LCO\ ({\it bottom}). All the plots span 2.5 dex vertically for direct comparison. In each panel, arrows indicate lower limits of the $x$-axis, and the solid line shows the relation computed using the {\it locfit.censor} function. The correlation coefficient and probability of null hypothesis are also given in the bottom panel for the \LCII/\LCO-$\Sigma_{\mathrm{IR}}$ relation.}
\label{Fig_sigfir}
\end{figure}
\subsubsection{Dependence of \LCII/\LCO\ on the IR Surface Brightness} 
As mentioned in \S3.3, the \LCII/\LFIR\ ratios of local ULIRGs may be smaller by as much as several $\times$10 compared to local less luminous galaxies. This ``\CII/FIR" is found to correlate well with the total infrared luminosity surface density (e.g., \citealt{2017ApJ...846...32D,2022ApJ...926...82S}), $\Sigma_{\mathrm{IR}}$, which consists of both the energy injection and the compactness of a region, thereby traces the heating rate in the ISM. Therefore, it is necessary to check how the ``\CII/FIR" deficit will affect \LCII/\LCO. 

For our main sample, only a part of the GOALS LIRG and local ULIRG subsample galaxies have IR size measurements in \cite{2016A&A...591A.136L}, and thus for these sources we plot \LCII/\LCO\ as a function of  $\Sigma_{\mathrm{IR}}$ ($\equiv \frac{L_{\mathrm{IR}}}{2 \pi r^2}$, where the radius $r$ is estimated from the 70\,\mum\ images) in the bottom panel of Figure \ref{Fig_sigfir}. Clearly, there exists a trend that \LCII/\LCO\ decreases as $\Sigma_{\mathrm{IR}}$ increases. To investigate this relation, we use the Kaplan–Meier estimate (\citealt{1958JASA...53...457K}) for censored data\footnote{Implemented in the {\it locfit.censor} function in the {\it locfit} package in {$\mathbf{R}$}.}. The resulting \LCII/\LCO$-\Sigma_{\mathrm{IR}}$, which has a steeper slope (about $-0.3$) when $\Sigma_{\mathrm{IR}}\gtrsim10^{11}$ $L_\odot$~kpc$^{-2}$, is shown by the solid line in the figure, with a scatter of 0.19 dex in \LCII/\LCO, reduced by $\sim$20\% compared with that (0.24 dex) of the original data. The trend might be caused by the following reasons: 

(1) The deficit of the \CII\ emission, for which there exist comprehensive studies (e.g., \citealt{2001ApJ...561..766M,2003ApJ...594..758L,2014PhR...541...45C,2015MNRAS.449.2883G,2016ApJ...824..146Z,2017ApJ...846...32D,2019ApJ...876..112R}) and several explanations (e.g., see \citealt{2014PhR...541...45C} for a review; \citealt{2016MNRAS.463.2085M,2017MNRAS.467...50N,2022ApJ...934..115B,2023MNRAS.tmp.3653L}) in the literature; and/or (2) a decreasing $\alpha_{\mathrm{CO}}$ with increasing gas surface density ($\Sigma_{\mathrm{gas}}$) when considering $\Sigma_{\mathrm{IR}} \propto \Sigma_{\mathrm {gas}}^N$ ($N\sim1.4$; e.g., see \citealt{1998ARA&A..36..189K}), as $L_{\mathrm{[C\,\scriptsize{\textsc{ii}}]}}/L_{\mathrm{CO}} \propto \alpha_{\mathrm{CO}}/\alpha_{\mathrm{[C\,\scriptsize{\textsc{ii}}]}}$. 

The latter point seems to be well consistent with the prescription of $\alpha_{\mathrm{CO}} \propto \Sigma_{\mathrm{gas}}^{-0.5} \sim \Sigma_{\mathrm{IR}}^{-0.35}$ for objects with similar metallicities as proposed in \cite{2013ARA&A..51..207B}. If this argument is true, it would suggest that \alphaCII\ almost does not vary with $\Sigma_{\mathrm{IR}}$, and the \CII\ emission can serve an excellent molecular gas mass tracer for extreme starburst systems. As discussed in \S\ref{sect:alphaco}, however, it appears likely that the current cross-calibration of molecular gas mass tracers in metal-rich galaxies (\citealt{2022MNRAS.517..962D}) does not favour a significantly variable $\alpha_{\mathrm{CO}}$. For example, Arp~220 and NGC~4418, two of the most compact ($\Sigma_{\mathrm{IR}}\sim10^{12.7}~L_\odot$~kpc$^{-2}$) starburst galaxies in our sample, have an averaged $\alpha_{\mathrm{CO}}$ of $3.6\pm0.5$ and $4.4\pm0.7$~$M_\odot$~(\kkmspc)$^{-1}$ (\citealt{2022MNRAS.517..962D}), respectively. 

Regarding the first point, we further explore it by plotting the \CII- and \COI-to-IR luminosity ratios as a function of $\Sigma_{\mathrm{IR}}$ in the top and middle panels of Figure \ref{Fig_sigfir}, respectively. Since all the three plots in Figure \ref{Fig_sigfir} span 2.5 dex vertically, we can make a direct comparison of the dependence on $\Sigma_{\mathrm{IR}}$ among different line-to-IR luminosity ratios. We can see that the \LCII/\LIR-$\Sigma_{\mathrm{IR}}$ relation shows a steeper slope than \LCO/\LIR-$\Sigma_{\mathrm{IR}}$ (see also \citealt{2020A&A...641A.155V} for a similar trend in \LCO/\LIR-$\Delta(\mathrm{MS})$ relation), resulting in the dependence of \LCII/\LCO\ on $\Sigma_{\mathrm{IR}}$. Therefore, the ``\CII/FIR" deficit is responsible for systematically decreasing in \LCII/\LCO\ with increasing $\Sigma_{\mathrm{IR}}$.

According to the relation shown in the bottom panel of Figure \ref{Fig_sigfir}, $M_{\mathrm{mol}}$ of those galaxies with $\Sigma_{\mathrm{IR}}~\gtrsim~10^{12}~L_\odot$~kpc$^{-2}$ is likely to be systematically underestimated by a factor of $\gtrsim$2 due to the deficit of the \CII\ emission. Furthermore, the total molecular gas mass can be underestimated by as much as about an order of magnitude for extreme cases like NGC~4418, though such systems are very rare (only 1 case) in our sample. In general, the dependence of \LCII/\LCO\ on $\Sigma_{\mathrm{IR}}$ only results in a small increase in the total scatter based on our sample.

\subsubsection{Relation between \LCII/\LCO\ and Distance off Main Sequence} 
\begin{figure}[t]
\centering
\includegraphics[width=0.46\textwidth,bb = 9 23 418 418]{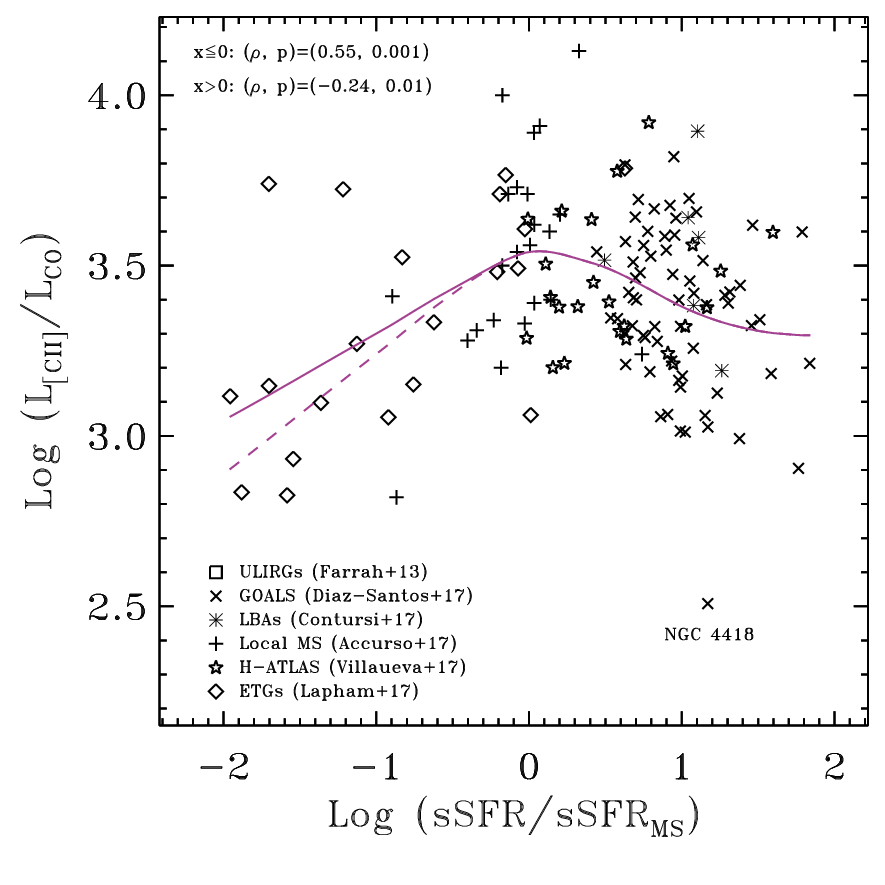}
\caption{\LCII/\LCO\ plotted against sSFR/sSFR$_{\mathrm{MS}}$. The solid purple line shows the relationship obtained with the $\it locfit$ function using all data points, while the dashed line displays the relationship estimated by excluding the two apparent outliers (i.e., two ETGs with $\Delta(\mathrm{MS}) < -1$ and $\log (L_{\mathrm{[C\,\scriptsize{\textsc{ii}}]}}/L_{\mathrm{CO}})>3.7$). The correlation coefficient and probability of null hypothesis, computed separately for $\log (\mathrm{sSFR/sSFR_{MS}}) \leq 0$ and $\log (\mathrm{sSFR/sSFR_{MS}}) > 0$, are also given in the plot.}
\label{Fig_ssfr}
\end{figure} 

Employing a Bayesian analysis, \cite{2017MNRAS.470.4750A} find that the distance from the main sequence ($\equiv\Delta(\mathrm{MS})$), which is related to the strength of UV radiation field, plays a secondary but statistically important role in determining the \LCII/\LCO\ ratio, in the sense that \LCII/\LCO\ monotonically increases with $\Delta(\mathrm{MS})$ within the parameter space constrained ($-0.8<\Delta(\mathrm{MS})<1.3$). Here the distance off the main sequence is defined as:
\begin{equation}
    \Delta(\mathrm{MS}) = \log \frac{\mathrm{sSFR}}{\mathrm{sSFR_{MS}}}
\end{equation}
where sSFR is the specific star formation rate (i.e., SFR/$M_\star$), and sSFR$_{\mathrm{MS}}$ the sSFR for main sequence galaxies, calculated with the same equation (adopted from \citealt{2012ApJ...754L..29W}) in \cite{2017MNRAS.470.4750A}. 
To further investigate the relation between \LCII/\LCO\ and $\Delta(\mathrm{MS})$ for a larger dynamic range, we plot \LCII/\LCO\ as a function of $\Delta(\mathrm{MS})$ in Figure \ref{Fig_ssfr}, for the 138 objects with stellar mass measurements. Same as what found in \cite{2017MNRAS.470.4750A}, indeed, \LCII/\LCO\ increases with $\Delta(\mathrm{MS})$ when $-2<\Delta(\mathrm{MS}) \leq 0$, as illustrated by the solid purple line obtained with the $\it locfit$ function in $\mathbf{R}$. The Spearman correlation coefficient is $\rho=0.55$, with the probability of null hypothesis $p=0.001$, indicating a moderate positive correlation. If the two apparent outliers (i.e., two ETGs with $\Delta(\mathrm{MS}) < -1$ and $\log (L_{\mathrm{[C\,\scriptsize{\textsc{ii}}]}}/L_{\mathrm{CO}})>3.7$) are excluded from the fitting, an even stronger correlation can be obtained (as shown by the dashed line), with $\rho=0.74$ and $p=1.69\times10^{-6}$. This result suggests that $M_{\mathrm{mol}}$ could be systematically underestimated by a factor of $2-3$ for galaxies with $\Delta(\mathrm{MS}) \lesssim -1.5$.

When $0<\Delta(\mathrm{MS})<2$, by contrast, there only exists a very weak, negative ($\rho=-0.24$ and $p=0.01$) correlation, indicating that \LCII/\LCO\ decreases slowly as $\Delta(\mathrm{MS})$ increases. The difference between our and \cite{2017MNRAS.470.4750A} results is believed to be caused mainly by the samples used and the sample sizes, as Accurso et al. sample comprises 30 metal-rich and metal-poor galaxies, only eight out of which have $\Delta(\mathrm{MS})~>~0.3$. Our results suggest that the \LCII/\LCO\ ratio is mainly determined by other physical parameter(s) for metal-rich, starburst systems.

\subsubsection{Effect of the Contribution from Ionized Gas to the \CII\ Emission}
\label{sec:ionizedcontribution}
In our analysis, the contribution from ionized gas to the \CII\ emission is not subtracted, and thus the molecular gas mass might be overestimated for the systems that the \CII\ emission is dominated by the ionized gas. To check whether this hypothesis is true, we plot the \LCII/\LCO\ ratio against the \LNII/\LCII\ ratio in the middle panel of Figure \ref{FigCII2CO_fagn}. As mentioned in \S\ref{sect:cii_origin}, \LNII/\LCII\ can be used to estimate the fractional contribution ($f_{\mathrm{[C\,\scriptsize{\textsc{ii}}],ion}}$) from the diffuse ionized gas to the total \CII\ flux, namely $f_{\mathrm{[C\,\scriptsize{\textsc{ii}}],ion}} = \frac{[\mathrm{C}\,\scriptsize{\textsc{ii}}]_{\mathrm{ion}}}{[\mathrm{C}\,\scriptsize{\textsc{ii}}]_{\mathrm{total}}} = \left (\frac{[\mathrm{C}\,\scriptsize{\textsc{ii}}]}{[\mathrm{N}\,\scriptsize{\textsc{ii}}]} \right)_{\mathrm{theo}} \times \left (\frac{[\mathrm{N}\,\scriptsize{\textsc{ii}}]}{[\mathrm{C}\,\scriptsize{\textsc{ii}}]} \right)_{\mathrm{obs}}$, where $ \left (\frac{[\mathrm{C}\,\scriptsize{\textsc{ii}}]}{[\mathrm{N}\,\scriptsize{\textsc{ii}}]} \right)_{\mathrm{theo}}$ is the predicted \LCII/\LNII\ ratio in diffuse ionized gas. This is because that, $\left (\frac{[\mathrm{C}\,\scriptsize{\textsc{ii}}]}{[\mathrm{N}\,\scriptsize{\textsc{ii}}]} \right)_{\mathrm{theo}}$ only shows a very weak dependence on electron density and temperature, and thus is almost constant for a given carbon-to-nitrogen abundance ratio (C/N; see, e.g., \citealt{2006ApJ...652L.125O}; \citealt{2018MNRAS.481...59Z}). We note that the C/N ratio in galaxies is nearly independent on metallicity, showing a dispersion of $\sigma=0.20$ dex (\citealt{2019ApJ...874...93B}).  

As demonstrated in the middle panel of Figure \ref{FigCII2CO_fagn}, there exist no obvious trend between these two ratios. The Kendall’s $\tau$ correlation coefficient, computed using the {\it cenken} function in the {\it NADA} package within {\it $\mathbf{R}$}, is $\tau = -0.35$, with a $p$-value of $3.2\times10^{-7}$, indicating a weak anti-correlation between \LNII/\LCII\ and \LCII/\LCO. This anti-correlation suggests that a larger $f_{\mathrm{[C\,\scriptsize{\textsc{ii}}],ion}}$ will lead to a lower \LCII/\LCO\ ratio, opposite to the aforementioned hypothesis that the contribution of ionized gas to the \CII\ emission would overestimate the total molecular gas mass. This result appears to be understandable in the sense that the lower the SF activity level is (as a consequence, a smaller $G_0/n$ and thus lower \LCII/\LCO), the higher the fractional contribution from the ionized gas is likely to happen (e.g., 
\citealt{2017ApJ...846...32D}).


\section{Summary}
In this paper we present the relation between the \CII\ line and the \COI\ emission for a sample comprising about 200 local and high-$z$ galaxies, including ETGs, main-sequence galaxies, (U)LIRGs/SMGs, and QSOs/AGNs. We explore the origin of this correlation, as well as investigate the dependence of the \CII-to-\COI\ luminosity ratio on different physical properties such as the IR luminosity surface density ($\Sigma_{\mathrm{IR}}$), hardness of the UV radiation field (\OIIIFIR/\NIIbc), distance to the main sequence ($\Delta(\mathrm{MS})$) and the fractional contribution of ionized gas (\NIIab/\CII). We show that the \CII\ emission can generally serve as a good tracer of the total molecular gas mass with an uncertainty of a factor of $\lesssim$2. However, caution needs to be taken when applying a constant \CII-to-$M_{\mathrm{H_2}}$ conversion factor to estimate the molecular gas content in some extreme cases. For galaxies having low-level SF activities (i.e., $\Delta(\mathrm{MS}) \lesssim -1.5$) or high SFR surface density (i.e., $\Sigma_{\mathrm{IR}}\gtrsim10^{12}~L_\odot$~kpc$^{-2}$), $M_{\mathrm{mol}}$ may be systematically underestimated by a factor of $\gtrsim$2. Our main results are:
\begin{enumerate}
\item There exists a strong and tight (rms scatter $\lesssim$0.3 dex), linear correlation between \LCII\ and \LCO\ in massive galaxies located at different redshifts, i.e., $\log L_{\mathrm{CO}} = (-3.39\pm0.21)+(1.00\pm0.02)\log L_{\mathrm{[C\,\scriptsize{\textsc{ii}}]}}$, confirming the \CII's capability to trace total molecular gas.

\item Based on the observed \CII/\COI\ ratios and $\alpha_{\mathrm{CO}}$ derived most recently (\citealt{2022MNRAS.517..962D}), we obtain the \CII-to-\MHH\ conversion factor, $\alpha_{[\mathrm{C\,\scriptsize{\textsc{ii}}}]} = 34.9$ $M_\odot/L_\odot$ with an uncertainty of $\sigma=0.26$~dex. There only exists a small difference between (U)LIRGs ($\alpha_{[\mathrm{C\,\scriptsize{\textsc{ii}}}]} = 34.0$ $M_\odot/L_\odot$, and $\sigma=0.25$~dex) and less-luminous galaxies ($\alpha_{[\mathrm{C\,\scriptsize{\textsc{ii}}}]} = 32.5$ $M_\odot/L_\odot$ and $\sigma=0.33$~dex) if they are analyzed separately. 

\item By comparing the distributions between observed and modelled \CII/\COI\ ratios, it is found that the tight and linear relation between \CII\ and \COI\ is very likely determined by the range of $G_0/n$ and average $A_V$ in galaxies.  

\item The \CII/\COI\ ratio anti-correlates with $\Sigma_{\mathrm{IR}}$ for the subsample of local (U)LIRGs, and the relation becomes steeper when $\Sigma_{\mathrm{IR}}\gtrsim10^{11}~L_\odot\,\mathrm{kpc}^{-2}$. The anti-correlation might be ascribed to the well-known deficit of \CII\ in such kind of galaxies.

\item The \CII/\COI\ ratio correlates positively with $\Delta(\mathrm{MS})$ when $\Delta(\mathrm{MS}) \lesssim 0$, and there is only a very weak, negative relation between these two parameters when $\Delta(\mathrm{MS})$ becomes higher.

\item For systems in which the \CII\ emission is dominated by ionized gas, the \CII/\COI\ ratio tends to show a systematically smaller value, resulting in an underestimation of the total molecular gas.

\end{enumerate}

\begin{acknowledgments}
We thank the anonymous referee for useful comments/suggestions that significantly improved the paper. YZ is grateful for support from the National Natural Science Foundation of China (NSFC) under grant No. 12173079. 
\end{acknowledgments}



\bibliography{My_AAs_arxiv.bib}
\bibliographystyle{aasjournal}



\end{document}